# Fill Factor Losses and Deviations from the Superposition Principle in Lead-Halide Perovskite Solar Cells

*David Grabowski, Zhifa Liu, Gunnar Schöpe, Uwe Rau and Thomas Kirchartz\**

D. Grabowski, Z. Liu, G. Schöpe, U. Rau
IEK5-Photovoltaik, Forschungszentrum Jülich, 52425 Jülich, Germany

T. Kirchartz
IEK5-Photovoltaik, Forschungszentrum Jülich, 52425 Jülich, Germany
Faculty of Engineering and CENIDE, University of Duisburg-Essen, Carl-Benz-Str. 199, 47057 Duisburg, Germany
E-mail: t.kirchartz@fz-juelich.de

**Abstract:**

The enhancement of the fill factor in the current generation of perovskite solar cells is the key for further efficiency improvement. Thus, methods to quantify the fill factor losses are urgently needed. A classical method to quantify Ohmic and non-Ohmic resistive losses in solar cells is based on the comparison between the voltage in the dark and under illumination analysed at equal recombination current density. Applied to perovskite solar cells, we observe a combination of an Ohmic series resistance with a voltage-dependent resistance that is most prominent at short circuit and low forward bias. The latter is most likely caused by the poor transport properties of the electron and/or hole transport layers. By measuring the photoluminescence of perovskite solar cells as a function of applied voltage, we provide direct evidence for a high quasi-Fermi level splitting at low and moderate forward bias that substantially exceeds the externally applied voltage. This quasi-Fermi level splitting causes recombination losses and, thus, reduces both the short-circuit current and the fill factor of the solar cell.

# 1. Introduction

With reports of > 25% efficiency,[1-4] lead-halide perovskite solar cells continue to close the efficiency gap to established photovoltaic technologies such as crystalline Si (26.7%)[5, 6] or GaAs (29.1%). Even more, tandem solar cells using perovskite subcells have now achieved efficiencies close to 30%[7] thus, beating even the best single junction solar cells. The achieved efficiencies are especially remarkable given the fact that we compare polycrystalline thin-film absorbers with monocrystalline materials with decades of development. While the key obstacles towards commercialization are issues of stability, scalability and toxicity of materials and solvents,[8] it is still important to study the remaining efficiency losses of halide perovskite solar cells with the help of reliable characterization methods. As shown recently in a meta-study[9] mainly dedicated to non-radiative recombination losses, the fill factor is one of the key remaining problems to achieve even higher efficiencies and to close the gap to efficiencies > 30% allowed by the Shockley-Queisser (SQ) model[10] for a single junction solar cell.

The present paper discusses the influence of resisive losses on the fill factor of metal-halide perovskite solar cells starting with a literature overview (section 2). Departures from the maximum possible fill factor are then analysed in section 3 by light intensity dependent current-voltage curves.[11-13] Here, we show how they can be used to determine external (voltage independent) and internal (voltage dependent) contributions to the series resistance of perovskite solar cells. By using numerical simulations in section 4, we then show how different parameters such as the mobilities of absorber and contact layers affect the internal and external series resistance. Finally, section 5 uses voltage dependent photoluminescence to explain the apparent shunt of the illuminated current-voltage curve and its relation to the poor collection of photogenerated free charge carriers. By combining current-voltage curves with photoluminescence voltage curves, we can also provide an estimate of the recombination losses as a function of the external voltage.

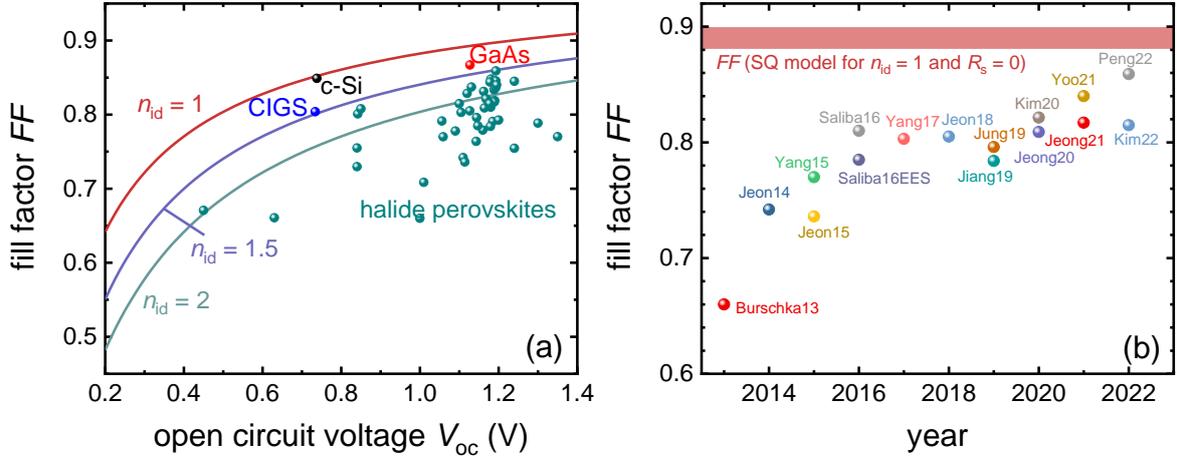

**Figure 1**: (a) Fill factor as a function of open-circuit voltage according to equation (1) for $n_{id}$ = 1, 1.5 and 2 (lines) compared to experimental data points for halide perovskites[14] compared to relatively mature technologies such as crystalline Si (c-Si), GaAs and Cu(In,Ga)Se$_2$ (CIGS). (b) Fill factor as a function of time for the same set of halide perovskite solar cells as shown in panel (a). Each of the data points represents a state of the art efficiency at the time of publication.[1-4, 15-26]

## 2. Status of fill factor losses in perovskite solar cells

**Figure 1** shows the fill factor in the SQ model as a function of the open-circuit voltage of the cell in comparison with the phenomenological equation from Refs. [27, 28]

$$FF_0 = \frac{v_{oc} - \ln(v_{oc} + 0.72)}{v_{oc} + 1}. \quad (1)$$

where the normalized open circuit voltage is defined as $v_{oc} = qV_{oc}/(n_{id}k_BT)$. Note that the *FF* only depends on $V_{oc}$, the temperature *T* (via the thermal voltage $kT/q$) and the ideality factor $n_{id}$ and does not consider any resistive losses.

In the SQ model, we have $n_{id}$ = 1, i.e. the red line in Figure 1a corresponds to this ideal result while the blue and green lines represent less ideal cases for $n_{id}$ = 1.5 and 2, corresponding to less ideal recombination mechanisms. The symbols in Figure 1a represent data from lead-halide perovskite solar cells (green) that were state of the art at the time of publication or represent a record efficiency at a certain band gap according to the current version of the emerging PV reports.[14] Those values are compared to the best crystalline Si (c-Si), GaAs and Cu(In,Ga)Se$_2$ (CIGS) solar cells.[29] The data points of these three comparably mature photovoltaic technologies approach the ideal case (for Si and GaAs) or lie close to the $n_{id}$ = 1.5 line (CIGS).

Until recently, the fill factors of halide perovskite solar cells were not exceeding the $n_{id}$ = 2 line. Within the last two years, however, new results[3, 26] have been reported that mean that the best perovskites approach the $n_{id}$ = 1.5 line and are therefore on a similar level as CIGS. Note that this doesn't necessarily mean that perovskite solar cells generally have a large $n_{id}$. The ideality factor is only infrequently reported thereby making meta-analyses on the topic difficult. A few notable exceptions are refs. [7, 30-35] that report a wide range of ideality factors between close to one to around two. One recent example where data is available is the perovskite solar cell that shared most of the device geometry with top cell of the 29% tandem in ref. [7]. This single junction solar cell had an ideality factor of 1.26, i.e. much lower than 2, and an open circuit voltage $V_{oc}$ = 1.15 V. Given that the fill factor was 84.0% (exceptionally high for perovskite solar cell) and $FF_0(n_{id}=1.26, V_{oc} = 1.15$ V$)$ = 87.4 % there must have been a substantial resistive contribution to the fill factor in these cells that is not explicitly included in Figure 1a. Figure 1b shows the trend in $FF$ over time using cells with highly competitive or even record efficiencies at the time of publication. We note that while there was progress at the beginning, this progress stagnated for several years before some results in the last two years show that there is the potential to substantially overcome the 80% barrier.

While the fill factor is therefore key to further efficiency improvements, it is not common in the community to report key performance parameters determining the fill factor. The classical description of the fill factor of *pn*-junction solar cells would predict that there are four key criteria that determine the fill factor. The first and second parameters are the before-mentioned $V_{oc}$ and $n_{id}$ that determine $FF_0$. The other two parameters are the series resistance $R_s$ and the parallel or shunt resistance $R_p$. Usually, the open-circuit voltage within one technology varies only by a small amount. Hence, the effect of $V_{oc}$ on $FF$ is important mostly for comparison between different technologies. In highly efficient and non-degraded solar cells, usually the parallel resistance extracted from the dark $JV$ curve is high enough to not substantially reduce the fill factor. This leaves the ideality factor and the series resistance as the two key remaining parameters that have to be measured and understood for further device optimization. For Si solar cells as an example, there are a range of methods available[36, 37] to determine the series resistance and detailed analyses are common[38] in order to determine the different contributions to $R_s$. In thin-film solar cells, however, the situation is often more complicated. While there are still substantial resistive effects, not all of them are necessarily of an Ohmic nature. In particular, resistive effects originating from undoped and highly resistive parts of the absorber itself often contribute substantially to a non-Ohmic series resistance. The fact that the series resistance becomes non-Ohmic results from the fact that the conductivity of any undoped semiconductor layers in the stack depends on carrier density and thereby on voltage and illumination intensity. Note that this is not the case for sufficiently doped semiconductors, where the majority carrier density (and thereby the conductivity) does not change with voltage and illumination up to a certain point, where the semiconductor undergoes a transition from low-level to high-level injection. Thus, if the transition happens outside the range of interest (i.e. one sun maximum power point), the series resistance would be fairly constant as a function of voltage and illuminaton. The fact that this is not the case for most thin-film solar cells has two consequences, namely (i) that the superposition principle[39-41] often stops to be valid in thin-film solar cells and (ii) that the series resistance is no longer a constant but a function of voltage and illumination.[12] These two consequences, are certainly part of the reason, why the community did not yet adapt clear standards how to measure and report parameters such as $R_s$ and $n_{id}$ that would support our understanding of fill factor losses.

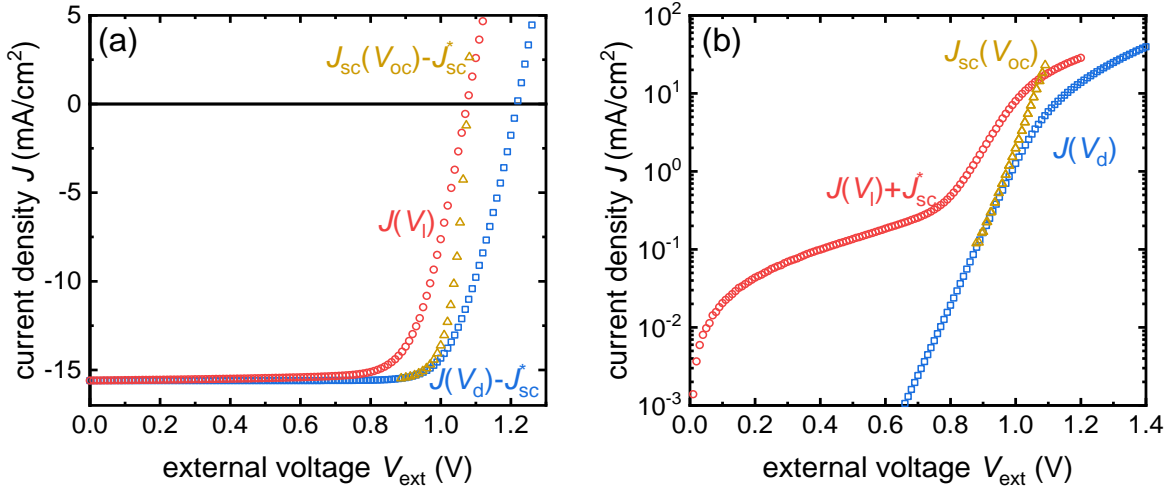

**Figure 2:** (a) Linear and (b) semilogarithmic plot of the current-voltage characteristic of a coevaporated MAPI solar cell in the dark $J(V_d)$ (blue) and under one sun illumination $J(V_l)$ (red) as well as the $J_{sc}$-$V_{oc}$ curve (yellow). The curves for $J(V_d)$ and $J_{sc}$-$V_{oc}$ were moved to the (a) fourth quadrant or (b) first quadrant of the coordinate system by subtraction (or addition) of the short-circuit current density $J_{sc}^*$ taken from $J(V_l)$ as indicated in the figure. See Figure S5 for an equivalent graph for a solution-processed MAPI solar cell.

## 3. Voltage Dependent Series Resistance from Current-Voltage Measurements

The superposition principle[39] is one of the most frequently taught concepts, when explaining the current-voltage curve of a solar cell in textbooks and lectures. Typically, the current voltage curve of an idealized solar cell is written as

$$J_l = J_0 \left( \exp\left(\frac{qV_l}{n_{id}kT}\right) - 1 \right) - J_{sc}^* = J_d - J_{sc}^* \qquad (2)$$

where $J_0$ is the saturation current density of the solar cell. Here, we use the index l (for light) to highlight that we refer to the current density $J_l$ and external voltage $V_l$ of the illuminated (one-sun) current voltage curve and the index d (for dark) to indicate the dark current. Equation 2 claims that the current denstiy of a solar cell under illumination is the same as the one in the dark minus a short circuit current density $J_{sc}^*$ that depends on illumination but not on voltage. Given that the dark current density of an ideal diode is a recombination current density, equation 2 can also be interpreted as saying that recombination only depends on voltage but not on voltage and illumination. Thus, if we were to apply 0.5 V forward bias to the solar cell, there would be a certain recombination flux within the solar cell that would not depend on the illumination intensity as long as we keep the 0.5 V constant. The concept of equation 2 is known in photovoltaics as the superposition principle and in the literature on photovoltaics two important types of deviations from the superposition principle have been discussed.[39, 41] The first deviation is the situation, where recombination at a given voltage does depend on illumination.[41-43] This deviation is always present in a solar cell but it is often (but not always) negligible, when analyzing the limits to the efficiency. A simple way of quantifying this deviation is to measure the carrier density at short circuit under illumination for instance by photoluminescence[44-47], photoconductance,[48] photocapacitance[49, 50] or charge extraction techniques.[51, 52] At around one sun illumination, photoluminescence at short circuit is essentially measurable in every solar cell technology and thereby provides evidence for a non-neglible concentration of non-extracted electrons and holes. The second deviation is the presence of a series resistance $R_s$. This deviation is also always present and can be exploited to determine the value of $R_s$.

The first version of the method applied here was introduced as early as 1963 by Wolf and Rauschenbach[53] and was later refined by others.[11, 12] The method involves the comparison of dark and illuminated current-voltage characteristics. In **figure 2a**, we see a dark current-voltage curve in comparison to a current voltage curve under one-sun illumination as well as a $J_{sc}$-$V_{oc}$ curve generated by measuring the short-circuit current density $J_{sc}$ and the open-circuit voltage $V_{oc}$ at varied illumination intensities. The dark $JV$ curve as well as the $J_{sc}$-$V_{oc}$ curve were shifted into the fourth quadrant by subtracting the $J_{sc}^*$ of the illuminated current-voltage curve at one sun illumination. In **figure 2b**, we see the same data in the first quadrant using a logarithmic current axis. Here, the dark $JV$ curve as well as the $J_{sc}$-$V_{oc}$ curve were left untouched and the illuminated $JV$ curve was shifted up by $J_{sc}^*$.

The data shown in Figure 2 are taken from a coevaporated methylammonium-lead-iodide perovskite solar cell with the layer stack glass/ITO/PEDOT:PSS/poly-TPD/CH$_3$NH$_3$PbI$_3$ (MAPI)/PCBM/BCP/Ag. Here, PEDOT:PSS is poly(3,4-ethylenedioxythiophene) polystyrene sulfonate and poly-TPD is Poly(4-butylphenyldiphenyl-amine), MA is methylammonium (CH$_3$NH$_3$), PCBM is [6,6]-phenyl-C$_{61}$-butyric acid methyl ester and BCP is bathocuproine. The solar cell serves here as a rather arbitrary example of a non-ideal perovskite solar cell that allows us to illustrate the concept of the method. Figure S5 in the supporting information shows the same plot for an example of a solution-processed MAPI solar cell for comparison.

Upon inspection of the curves in Figure 2a, we see that the curves share (per definition) the same $J_{sc}^*$. However, their behaviour at high external voltages $V_{ext}$ differs quite substantially. This difference is partly caused by the series resistance $R_s$ of the cell. Furthermore, we note that in Figure 2b, the illuminated current voltage (shifted up by $J_{sc}^*$) shows a feature that resembles the effect of a low shunt resistance but that is not visible in the dark $JV$ curve. We will refer to this feature as a 'photoshunt', but note that its origin is very different from that of a shunt in the dark. We will now first focus on the effects of the series resistance and then later return to the explanation of the photoshunt.

In order to better understand the effect of $R_s$ on the three current-voltage curves, let us briefly recapitulate the one-diode equivalent-circuit model of a solar cell with a non-zero series resistance. It predicts that the current voltage curve under illumination is given by

$$J_l = J_0 \left( \exp\left( \frac{q(V_l - J_l R_s)}{n_{id} kT} \right) - 1 \right) - J_{sc}^* \qquad (3)$$

In the case of the shifted dark $JV$ curve, the current density is given by

$$J_d = J_0 \left( \exp\left( \frac{q(V_d - J_d R_s)}{n_{id} kT} \right) - 1 \right) - J_{sc}^* . \qquad (4)$$

Equations (3) and (4) are nearly identical except for the voltage drop over the series resistance which is $J_l R_s$ in equation (2) and $J_d R_s$ in equation (3) which differ only in the indices. If we compare the voltage at a given current density $J_l = J_d - J_{sc}^*$, we observe that the difference in the voltage drop over the series resistance is always the same and equal to $J_{sc}^* R_s$. Thus, for a perfect Ohmic, i.e. voltage independent, series resistance, the series resistance can be determined from the difference between $V_d$ and $V_l$ via

$$R_s = (V_d - V_l)/J_{sc}^* . \tag{5}$$

The easiest way to rationalize this difference in voltage drop is to imagine the situation at open circuit under illumination where $J_l = 0$ which must be compared to the situation in the dark where a forward current of $J_{sc}^*$ flows through the diode. For the dark current to have this magnitude, diode theory would predict that one needs a forward voltage that exceeds $V_{oc}$ under one sun conditions, by exactly $J_{sc}^* R_s$.

The blue symbols in **Figure 3** show the result of applying equation (5) to the data shown in Figure 2. The data shows a substantial voltage-independent series resistance visible at high voltages. This voltage independent resistance, we will denote as $R_{s,ext}$ in the following. Here it is approximately 6 to 10 Ωcm². In order to estimate the influence of a (constant) series resistance on the fill factor for a solar cell, it is useful to apply the approximate relations given by Green[27] that predict

$$\frac{\Delta FF}{FF_0} = \frac{R_s J_{sc}^*}{V_{oc}} \tag{6}$$

where $\Delta FF$ is the absolute loss in $FF$ due to the series resistance. If we assume typical values of $J_{sc} = 20$ mA/cm² and $V_{oc} = 1.1$V, then a resistance of 6 to 10 Ωcm² would correspond to a (quite substantial) relative fill factor loss of around 11% to 18%.

In addition to the comparison of the dark and light JV curve, it is also possible to involve the suns-$V_{oc}$ or $J_{sc}/V_{oc}$-curve. These are based on a measurement of $J_{sc}$ and $V_{oc}$ as a function of light intensity and provide a series-resistance-free version of the diode equation given by

$$J_{sc} = J_0 \left(\exp\left(\frac{qV_{oc}}{n_{id}kT}\right) - 1\right). \tag{7}$$

The $J_{sc}$/$V_{oc}$-curve is particularly useful to determine the ideality factor via a fit of ln($J_{sc}$) via $V_{oc}$ or by calculating the derivative[54]

$$n_{id} = \left(\frac{kT}{q}\frac{d\ln(J_{sc})}{dV_{oc}}\right)^{-1}. \tag{8}$$

By comparing the $J_{sc}$/$V_{oc}$-curve to the dark or light $JV$ curves, the series resistance can be calculated via

$$R_s = \frac{V_d - V_{oc}}{J_{sc}} \tag{9}$$

or

$$R_s = \frac{V_{oc} - V_l}{J_{sc}^* - J_{sc}}. \tag{10}$$

Thus, using equations (5), (9) and (10), there are in total 3 ways of determining the series resistance. Based on the simple idea of the equivalent circuit with a constant series resistance $R_s$, the three resulting curves should all lie on top of each other and be constant as a function of voltage or current at which the equations are evaluated. However, the perovskite solar cell data from Figure 2 does not behave in exactly the same way as a simple equivalent circuit would predict. This is partly due to physical reasons and partly due to experimental challenges in measuring small differences between quantities. At low voltages, the information obtained from equations (5) and (10) are nearly identical (blue and red data in Figure 3), because there is little difference between the dark $JV$ curve and the $J_{sc}$/$V_{oc}$-curve. This is due to the fact that i.e. around the maximum power point in Figure 2a, the resistive loss in the dark is minimal ($J_d = J_{sc}^* - J_{mpp}$ is small and so is $J_d R_s$), the loss in the $J_{sc}$/$V_{oc}$-curve is zero and the loss under illumination is substantial, i.e. $\sim J_{mpp} R_s$. For the same reason, the difference between $V_d$ and $V_{oc}$ is in practice nearly impossible to exploit, because the difference becomes very small. Especially in the context of small variations of the voltages due to hysteresis effects, analysing those small voltage differences to learn more about the series resistance is impossible. In the data shown in Figure 3, the apparent series resistance determined using equation (9) (green spheres) becomes negative for low voltages (below ~ 0.9V). Note that in Figure 3 as well as in the following, the voltage on the x-axis is always the higher of the two voltages whose difference is determined in equations 5, 9 and 10.

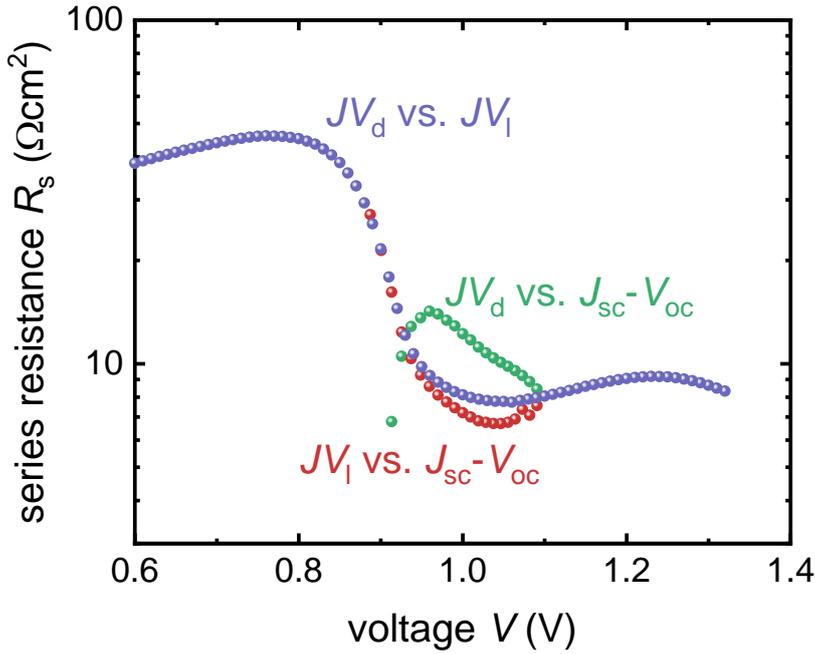

**Figure 3:** Series resistance $R_s$ as a function of externally applied voltage $V$ as determined from the comparison of different current-voltage characteristics. The $J_{sc}$-$V_{oc}$ curve in comparison to the dark current-voltage characteristic $J(V_d)$ (green) as well as the illuminated current-voltage curve $J(V_l)$ (red) and a comparison of the dark and the illuminated current-voltage characteristic (blue).

## 4. Simulating the Voltage-Dependent Series Resistance

In the previous section we have seen that the voltage-dependent determination of the series resistance $R_s(V)$ enables us to distinguish between Ohmic and non-Ohmic contributions. As a next step we are going to examine how certain parameters of the layers within a solar cell stack influence both the Ohmic contribution ($R_{s,ext}$) and the non-Ohmic contribution ($R_{s,int}$) to the series resistance. We will do this by employing numerical drift-diffusion simulations using the software ASA. The parameters for the simulation are given in Table I. The software ASA[55, 56] numerically solves the continuity equations for electrons and holes as well as the Poisson equation that links the space charge with the electrostatic potential. Transport of ionic charge is not explicitly considered in ASA, but we will mimic its effect in some of the calculations in this paper by using high permittivities in the perovskite layer that screen the electric field inside the perovskite layer.

**Table I:** Default parameter values used for the simulations if not otherwise stated.

| Parameter | PTAA | Perovskite | PCBM |
|---|---|---|---|
| Band gap $E_g$ | 3 eV | 1.55 eV | 1.8 eV |
| Electron affinity | 2.45 eV | 4 eV | 4.1 eV |
| Eff. DOS $N_C = N_V$ | $2 \times 10^{18}$ cm$^{-3}$ | $2 \times 10^{18}$ cm$^{-3}$ | $2 \times 10^{19}$ cm$^{-3}$ |
| Mobility $\mu_n = \mu_p$ | $10^{-3}$ cm$^2$/(Vs) | $10^2$ cm$^2$/(Vs) | $10^{-3}$ cm$^2$/(Vs) |
| Permittivity $\varepsilon_r$ | 3 | 30 | 3 |
| Interf. recom. velocity | 2 cm/s | | 2 cm/s |
| **General parameters** | | | |
| $R_s$ | | 6 Ωcm$^2$ | |
| $R_p$ | | $5 \times 10^3$ Ωcm$^2$ | |

**A. External Series Resistance**

Figure 4a shows the series resistance calculated using equations 5 and 10 in shades of violet and orange respectively for different values of the external series resistance. The external series resistance was varied logarithmically from 1 to 10 Ωcm$^2$ as indicated by the arrow. The series resistance according to equation 9 was omitted due to the practical problems of measuring it at low voltages. The effect of the external series resistance is visible especially at high voltages, where the total series resistance in the simulation saturates to a value that is slightly higher than the external series resistance. Thus, we note that there is a small nearly Ohmic contribution originating from those parts of the solar cell (i.e. absorber, ETL and HTL) that are explicitly simulated by ASA. Figure 4b shows the current-voltage curves that were used to obtain the series resistance shown in Figure 4a. Here, the solid lines represent the current-voltage curve under illumination shifted into the first quadrant by adding the respective $J_{sc}$, while the dotted lines are the dark current-voltage curves. Voltage differences between dotted and solid line of equal color are used to determine the series resistance according to equation (5). We observe that the shifted current voltage curves in Figure 4b show some photoshunt that is however in all cases much less dominant than the one seen in the experimental data shown in Figure 2b.

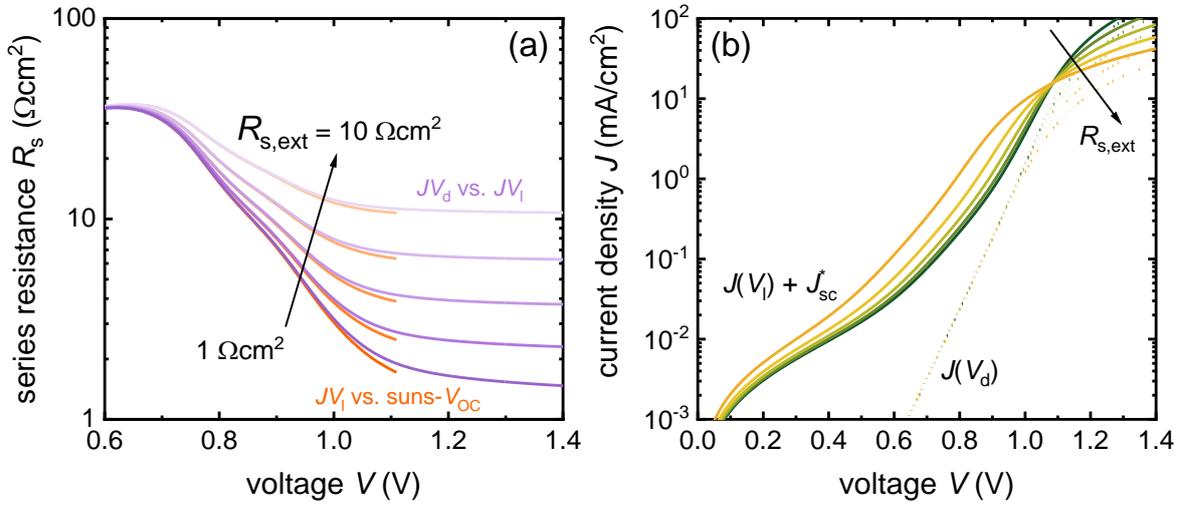

**Figure 4** (a) Simulated series resistance $R_s$ for varying values of the external series resistance $R_{s,ext}$ =1 – 10 Ωcm² implemented in the simulation. The series resistance was calculated using two different approaches comparing two current-voltage characteristics of the solar cell as denoted in the figure. (b) Illuminated current-voltage curve $J(V_l)$ shifted into the first quadrant by adding the respective $J_{sc}$ (solid line) and the dark $JV$ curve (dotted line). All curves are shown for the same variation of $R_{s,ext}$, arrows indicating the same directional change of the parameter as indicated in panel (a).

## B. Charge Carrier Mobilities of Transport Layers

Having discussed an external series resistance in the previous section, we will now turn to the effect of a series resistance originating from one of the transport layers. In this case, we choose the PCBM-based electron transport layer as an example and increase its resistance by reducing the electron mobility within the PCBM. Figure 5 shows the series resistances (a) and (c) as well as the current densities (b) and (d) for a simulated perovskite solar cell as a function of the mobility of the PCBM-based electron transport layer. The difference between the upper row of panels (a) and (b) versus the lower row of panels (c) and (d) is the permittivity of the perovskite layer. The upper row uses a relative permittivity of 30 which corresponds to the typically measured values.[57] From Figure 5a, we learn that a reduction in the PCBM mobility leads to an increased and increasingly voltage dependent series resistance that approaches a constant value at low voltages. This approximately constant value is caused by the fact that both $R_s = (V_d - V_l)/J_{sc}^*$ and $R_s = (V_{oc} - V_l)/(J_{sc}^* - J_{sc})$ will approach $V_d/J_{sc}^* \approx V_{oc}/J_{sc}^*$ for lower voltages, where $V_l \approx 0$ and $J_{sc}^* \gg J_{sc}$. Thus, once the apparent photoshunt seen in Figure 2b (red curve) leads to a rapid reduction of $V_l$, while $V_d$ and $V_{oc}$ stay high, the series resistances calculated using eqs. (4) and (9) become approximately constant or even decrease slightly.

In figure 5c and d, we used a very high permittivity in the perovskite ($10^4$) to emulate[58, 59] the situation where the electric field is screened in the active layer due to ionic motion.[60, 61] We again observe a nearly constant value of the resistance at low voltages. However, the transition from low to high resistances now resembles more closely the experimental situation seen in Figure 3. The transition is very abrupt with the the voltage position of the step strongly depending on the mobility of the transport layer.

Figure 5b and d show the associated dark and illuminated current-voltage curves, whereby the latter are shifted into the first quadrant by adding the respective $J_{sc}$ as described previously. We note that in the case without field-screening (Figure 5b), the photoshunt is much less pronounced than in the situation with field screening (Figure 5d). Generally, the situation with field screening resembles more closely the situation encountered in experiment (compare with Figures 2b and 3). Thus, the combination of field screening with a low mobility in the PCBM layer creates a qualitatively very different signature in the shifted JV curves under illumination and subsequently also in the series resistance than the situation without field screening.

To further illustrate the meaning of the resistance shown in Figure 5c, Figure 6 shows four band diagrams simulated for the case of field screening (high permittivity in the perovskite). The first column (panels a and c) shows the situation for a higher mobility in the PCBM and the second column shows the situation for a lower mobility. The first row shows band diagrams in the dark and the second row shows the band diagrams under illumination. Within each column, the recombination current is identical and chosen such that the band diagrams under illumination represent the situation at the maximum power point.

Figure 6a shows the situation in the dark for a high mobility in the PCBM. Here, all Fermi levels are basically flat and the external voltage $qV_{\text{ext,d}}$ is practically identical to the splitting of the quasi-Fermi levels. Figure 6b shows the situation for the lower PCBM mobility where a similar current has to be injected through the PCBM. However, given that the electron current has to follow the relation $J_n = n\,\mu_n\,dE_{\text{fn}}/dx$, and considering that $\mu_n$ has decreased by two orders of magnitude relative to Figure 6a, the gradient $|dE_{\text{fn}}/dx|$ now has to be substantially higher than before as seen in the visibly graded $E_{\text{fn}}$ (blue dashed line) in the PCBM (yellow region) in Figure 6b. As a result, the $qV_{\text{ext,d}}$ is substantially higher as compared to the quasi-Fermi level splitting.

Under illumination, the electrons and holes generated inside the perovskite absorber have to be extracted through the two charge transport layers. In Figure 6c, even for the higher mobility in the PCBM, a visible gradient $|dE_{\text{fn}}/dx|$ is seen that has the opposite sign to the gradient seen in the dark. Now, the external voltage is smaller than the quasi-Fermi level splitting in the perovskite. This gradient $|dE_{\text{fn}}/dx|$ increases substantially when reducing the mobility in the PCBM as seen in Figure 6d. The external voltage differences between the upper and lower panel of one column then define the series resistance seen in Figure 5c via equation 5. Thus, the origin of the series resistance are the gradients $|dE_{\text{fn}}/dx|$ seen in the PCBM layer that change in extent with the electron mobility in the PCBM.

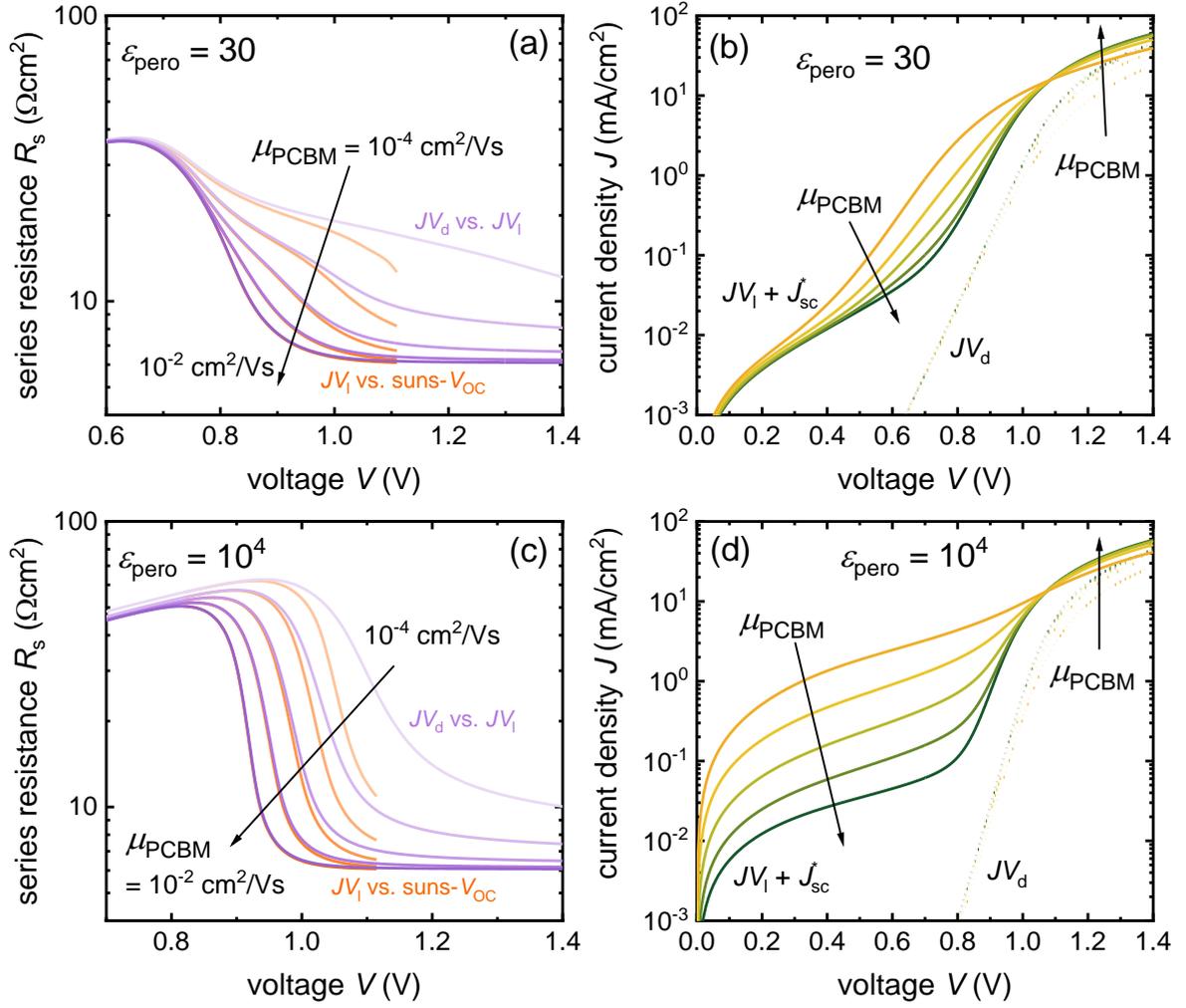

**Figure 5:** (a) and (b) Simulated series resistance $R_s$ for varying values of the external mobility of the PCBM layer ($\mu_{PCBM} = 10^{-4} - 10^{-2}$ cm²/Vs). The series resistance was calculated using two different approaches comparing two current-voltage characteristics of the solar cell as denoted in the figure. (b) and (d) Illuminated current-voltage curve $J(V_l)$ shifted into the first quadrant by adding the respective $J_{sc}$ (solid line) and the dark $JV$ curve (dotted line). Panels (a) and (b) show data for a relative permittivity of the absorber layer of 30, while panels (c) and (d) show the case for perfect field screening in the perovskite which was achieved by setting the permittivity to $10^4$. All curves are shown for the same variation of $R_{s,ext}$, arrows indicating the same directional change of the parameter as indicated in panel (a).

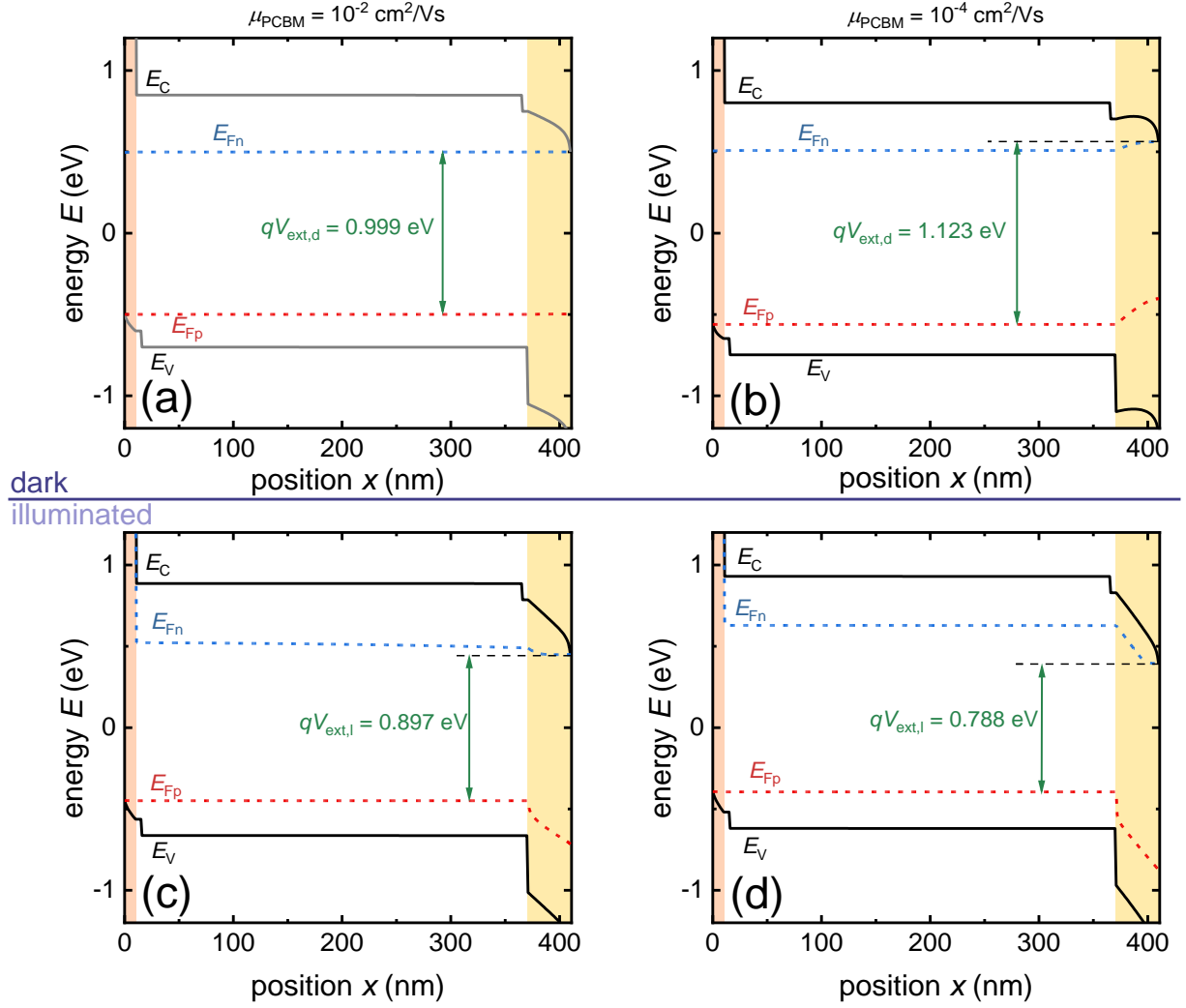

**Figure 6:** Simulated band diagrams of a perovskite solar cell with an electron mobility in the PCBM of (a, c) $\mu_{PCBM} = 10^{-2}$ cm$^2$/Vs and (b, d) $\mu_{PCBM} = 10^{-4}$ cm$^2$/Vs. The upper row of band diagrams represents the band diagrams in the dark at a current $J = J_{sc}-J_{mpp}$. The lower row are band diagrams under illumination at a current $-J_{mpp}$. Thus, the difference in voltage (between band diagrams in the same column) is the one used in equation (5) to calculate the series resistance.

## 5. Voltage Dependent Photoluminescence

It is obvious from the band diagrams in Figure 6 that the apparent photoshunt results from the difference between the Fermi-level splitting $\Delta E_F$ inside the perovskite absorber and the external voltage $V$ at the contacts (especially in Fig. 6d). An appropriate experiment to monitor both quantities, $\Delta E_F$ and $V$, simultaneously is the analysis of the voltage dependence of the photoluminescence.[45, 62] In most efficient perovskite solar cells, we can neglect strong Fermi level gradients at open circuit. Hence, we can assume at open circuit $\Delta E_F(V_{oc}) = qV_{oc}$. With this we can derive the Fermi level splitting $\Delta E_F(V)$ at each voltage via[63]

$$\Delta E_F(V) = kT\ln\left(\frac{\phi(V)}{\phi_{oc}}\right) + qV_{oc}, \qquad . \qquad (11)$$

where $\phi(V)$ is the photoluminescence intensity at a given voltage $V$ and $\phi_{oc}$ at open circuit.

To examine the voltage dependent photoluminescence, we used spin-coated perovskite solar cells with the layer stack glass/ITO/PTAA/MAPbI$_3$/PCBM/BCP/Ag, whereby PTAA is poly(triarylamine). The perovskite process uses a lead-acetate based precursor that leads to high open-circuit voltages > 1.2V[63, 64] and long charge carrier lifetimes > 20 µs.[65, 66] At the same time, the cells also suffer from only modest short-circuit currents and fill factors that suggest that charge extraction is non ideal.[63, 67] We then cut out the active area using a frequency tripled Nd:YVO$_4$ laser (355 nm wavelength, 6 ns pulse length, ~6 mW power). The rationale behind removing the perovskite directly adjacent to the active cell area is to avoid measuring any photoluminescence that could hit the detector and that originates from areas of the substrate that are at open circuit and not at the voltage defined by the source measure unit. Subsequently, we measured both the current voltage and the photoluminescence voltage curves at different illumination intensities using a λ = 532 nm laser diode. From the PL intensities, we then determined the $\Delta E_F(V)$ using equation 11.

Figure 7a shows the current voltage curves while Figure 7b shows the associated quasi-Fermi level splittings $\Delta E_F(V)$ measured using voltage-dependent photoluminescence and determined by Eq. (11). The most noteworthy observation in Figure 7b is the presence of luminescence $\phi_{SC} = \phi(0)$ at short circuit implying a considerable split of the quasi-Fermi levels $\Delta E_F^{SC} = \Delta E_F(0)$ which amounts, dependent on the illumination level, to values of around 1eV. Such high values of $\Delta E_F$ imply the presence of considerable recombination current densities $J_{rec}$ in the absorber such that the short circuit current density $J_{SC}$ is substantially reduced compared to the generation current density $J_{\text{gen}}$. The loss of short circuit current density $\Delta J_{\text{sc}} = J_{\text{rec}}(0) = J_{\text{gen}} - J_{\text{sc}}$ is related to the on-off ratio $\phi_{\text{sc}}/\phi_{\text{oc}}$ between the luminescence $\phi_{\text{sc}}$ at short circuit an $\phi_{\text{oc}}$ at open circuit via[45]

$$\frac{\Delta J_{\text{sc}}}{J_{\text{gen}}} = \left(\frac{\phi_{\text{sc}}}{\phi_{\text{oc}}}\right)^{1/n_{\text{id}}}. \tag{12}$$

where $n_{\text{id}}$ denotes the ideality factor of the recombination process. For the highest illumination intensity, $\Delta E_{\text{F}}(V_{\text{oc}}) = qV_{\text{oc}} = 1.23$ eV while $\Delta E_{\text{F}}(0) \approx 1.15$ eV, i.e. higher than the $qV_{\text{oc}}$ of many perovskite solar cells with a similar band gap. The difference between the Fermi level splitting at open and short circuit is about 80 meV corresponding to an on-off ratio $\phi_{\text{sc}}/\phi_{\text{oc}} = 0.05$. With an ideality factor $n_{\text{id}} = 2$ we obtain a relative loss $\Delta J_{\text{sc}}/J_{\text{gen}} = 0.2 = 20\%$. This is a significant loss that directly translates into an efficiency loss due to weak charge carrier collection. In comparison, for Cu(In,Ga)Se$_2$ solar cells, the measured on-off ratio was reported to be 1.7 % and the relative short circuit current loss was determined to be 4.3 % (ideality factor $n_{\text{id}} = 1.3$).[45]

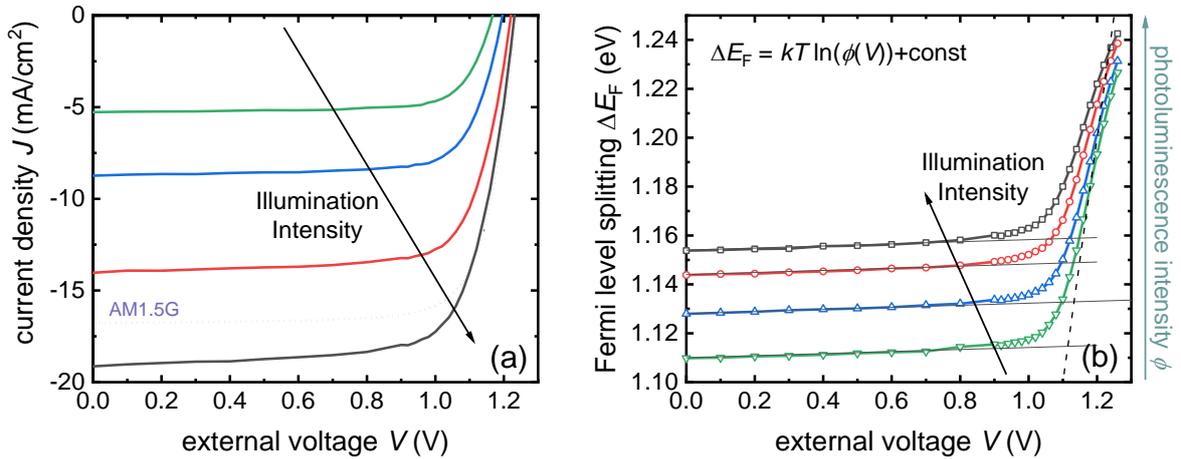

**Figure 7:** (a) Current voltage curves of a ~ 300 nm thick MAPbI$_3$ solar cell measured using a $\lambda = 532$ nm laser at different illumination intensities. The one-sun $JV$ curve (reverse scan) under a solar simulator is shown as dotted line. (b) For each illumination intensity, we measured the photoluminescence intensity in relative units and determined the quasi-Fermi level splitting $\Delta E_{\text{F}}(V)$ from the result using equation 11. This splitting was then plotted as a function of the external voltage showing that $\Delta E_{\text{F}}(V)$ is very high and rather voltage independent in the range of 0V to 1V.

The other notable observation in Figure 7b is the weak, almost linear, voltage dependence of $\Delta E_F(V)$ from short circuit until about 1 V. The slopes are about 0.004 eV/V for the range between 0 and 0.6 V independent of illumination intensity (see the thin solid lines in Fig. 7b).

## 6. Analysis of the Apparent Photoshunts

To better understand the apparent photoshunt that we have observed in the experimental data in Figure 2b and the various simulations shown in the previous section, it is instructive to study the voltage and illumination dependent recombination current in a perovskite solar cell. It is not a traditional shunt current that causes this behavior but instead it is an effect resulting from slow charge extraction at short circuit and low forward voltages. A reasonable assumption for the recombination current flowing in any solar cell at any bias condition would be to assume that it scales exponentially with the quasi-Fermi level splitting $\Delta E_F$. Depending on whether the cell is in high- or low-level injection and depending on which recombination mechanism is dominant, the exact dependence of the recombination current on $\Delta E_F$ may vary. However, we may use an ideality factor $n_{id}$ to describe any deviation from a proportionality of the form $\exp(\Delta E_F(V)/kT)$. If we call the prefactor for this relation $J_0$, we will therefore arrive at this form of the recombination current density

$$J_{rec} = J_0 \left( \exp\left(\frac{\Delta E_F(V)}{n_{id}kT}\right) \right). \tag{13}$$

The apparent photoshunt as plotted in Figure 2b (red curve) is then mathematically described by

$$J_{shift} = J_{rec}(V) - J_{rec}(0) = J_0 \left( \exp\left(\frac{\Delta E_F(V)}{n_{id}kT}\right) - \exp\left(\frac{\Delta E_F(0)}{n_{id}kT}\right) \right) \tag{14}$$

where $J_{shift}$ is the current density under illumination that is shifted up into the first quadrant by adding the short circuit current density. The resulting current density is zero at $V$=0 by definition and will increase when applying higher external voltages. In the absence of recombination, we would generate a photocurrent $J_{\text{gen}}$ at short circuit that would be given by $J_{\text{gen}} = J_0 \exp(qV_{\text{oc}}/(n_{\text{id}}kT))$. However, in the presence of recombination we would have to subtract the recombination current at short circuit. Thus, we could determine the prefactor $J_0$ of the recombination current density via

$$J_0 = \frac{J_{\text{sc}}}{\left(\exp\left(\frac{qV_{\text{oc}}}{n_{\text{id}}kT}\right) - \exp\left(\frac{\Delta E_F(0)}{n_{\text{id}}kT}\right)\right)} \tag{15}$$

and the total recombination current would consequently be given by

$$J_{\text{rec}} = \frac{J_{\text{sc}}\left(\exp\left(\frac{\Delta E_F(V)}{n_{\text{id}}kT}\right)\right)}{\left(\exp\left(\frac{qV_{\text{oc}}}{n_{\text{id}}kT}\right) - \exp\left(\frac{\Delta E_F(0)}{n_{\text{id}}kT}\right)\right)} = \frac{J_{\text{sc}}\phi^{\frac{1}{n_{\text{id}}}}(V)}{\phi_{\text{oc}}^{\frac{1}{n_{\text{id}}}} - \phi_{\text{sc}}^{\frac{1}{n_{\text{id}}}}}. \tag{16}$$

Thus, equation (16) allows us to calculate the recombination current only from the knowledge of $J_{\text{sc}}$, the ideality factor $n_{\text{id}}$ and the voltage dependent photoluminescence as shown in Figure 7b.

Figure 8a shows the current densities under illumination seen in Figure 7a shifted into the first quadrant by adding the respective short-circuit current density at the respective light intensity. We note that again the shifted currents look as if they were dark currents with a very low shunt resistance. The shifted current densities shown in Figure 8a, do not yet provide us with the actual recombination currents at each voltage, because they don't contain the information about the recombination current at short circuit. However, if we use equation 16, we obtain a voltage-dependent recombination current that quantifies the recombination losses at each voltage. We note that the recombination losses are substantial and reduce the $J_{\text{sc}}$ by several mA/cm² depending on the laser intensity. We also note that the exponential dependence of the recombination current on $\Delta E_F(V)$ could lead to large errors if the value of $\Delta E_F(V)$ was slightly off. However, the absolute values obtained are reasonable as the sum of recombination current and the measured $J_{\text{sc}}$ does not exceed the expected range for perfect collection. The $J_{\text{sc}}$ of the AM1.5G curve after the laser process to remove perovskite material adjacent to the cell area was 17 mA/cm² (average of forward and reverse scan). The predicted recombination current at short circuit (by interpolation between the red and black curve) is about 4.5 mA/cm², which would bring the total photocurrent in the absence of recombination to 21.5 mA/cm², which is certainly a conceivable value for a MAPI solar cell with a ~ 300 nm thick absorber layer.

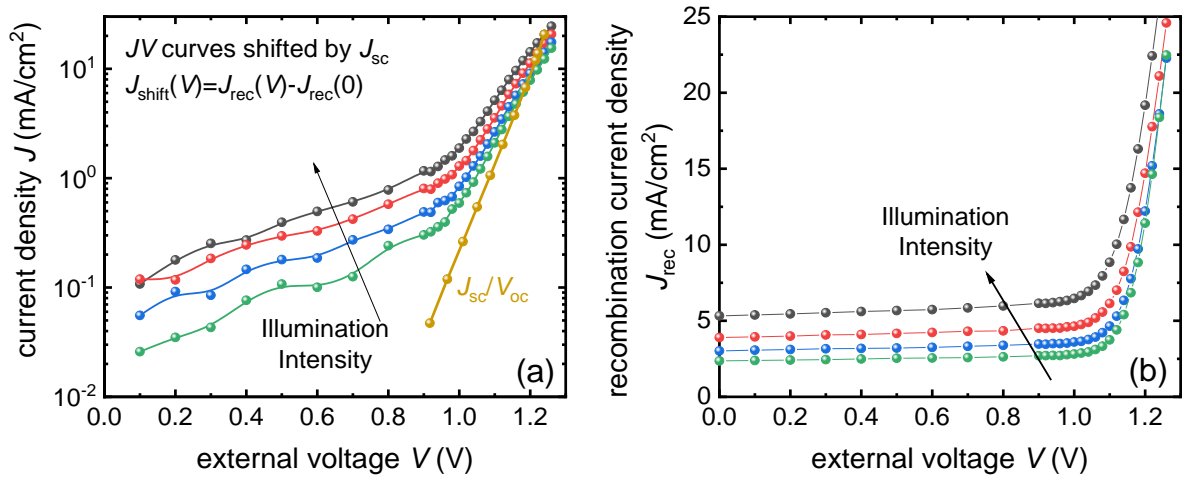

**Figure 8:** Current voltage curves shown in Figure 7a shifted into the first quadrant by adding the respective $J_{sc}$ for each curve. The $J_{sc}/V_{oc}$ curve is shown for comparison. (b) Recombination current along the whole $JV$ curve estimated from equation 16.

A key advantage of the present approach of quantifying the voltage dependent recombination current is the possibility to notice weakly voltage dependent charge collection losses. In other photovoltaic technologies such as organic solar cells and thin-film silicon solar cells, charge collection losses are of overwhelming importance. However, they are often relatively easy to identify, because of their strong voltage dependence.[68-71] Thus, in many cases the application of a reverse bias can substantially improve charge collection and therefore reveal the approximate magnitude of the losses.[70, 72] In halide perovskites, this is so far not a feasible approach because (i) the losses are only weakly voltage dependent and (ii) the application of a reverse bias often leads to reversible or irreversible changes in the device[73] thereby complicating any analysis.

## 7. Insights from the Shifted Currents $J_{shift}$

In the last section, we have shown that voltage-dependent photoluminescence is able to quantify the recombination current along the whole $JV$ curve and thereby allows us to also access the recombination current at short circuit. The downside of the approach is that it requires doing an additional measurement that may require using means (such as laser cutting) of avoiding the influence of signal from regions of the sample that are not contacted. The shifted current as shown e.g. in Figure 8a, however, is essentially free information that is available directly for every solar cell whose illuminated current-voltage curve has been measured. This thought naturally triggers the question whether we can obtain insights from an analysis of the shifted currents alone, i.e. without an additional photoluminescence measurement.

The simple fact that the shifted currents appear to have a shunt, i.e. an Ohmic contribution at low voltages provides some insights into the magnitude of the quasi-Fermi level change in the sample. As seen in Figure 9, this Ohmic region is indeed a region where current density is approximately linear with voltage. This can be seen best by checking the slope of the data in a double-logarithmic plot, where any linear function of the form $J \sim V$ has the slope $d\ln(J)/d\ln(V) = 1$. The simulated data originally shown in Figure 5d behaves quite similarly to the experimental data of the two different types of MAPI cells used in the paper so far. In both cases, the photoshunted region appears perfectly ohmic at low voltages and becomes slightly superlinear up to a certain voltage at which the curve shows a kink and the exponential part of the recombination current starts.

When looking at equation 14, we note that the equation describing $J_{shift}$ is not linear at all but composed of the difference of two exponentials. Thus, the question arises, how equation 14 produces this approximately linear behaviour. We can slightly rewrite equation 14 to obtain

$$J_{\text{shift}} = J_{\text{rec}}(0)\left(\exp\left(\frac{\Delta E_F(V)-\Delta E_F(0)}{n_{\text{id}}kT}\right) - 1\right). \quad (17)$$

which is a function of the form $e^x - 1$. Such a function has a Taylor expansion of the form $e^x - 1 = x + x^2/2 + \ldots$ for $x \to 0$. Thus, if the data appears linear vs. the external voltage, the higher order terms of the Taylor expansion must be small wrt. the linear term. Thus, we can deduce two things. First, $\Delta E_F(V) - \Delta E_F(0) \ll n_{\text{id}}kT$ must hold in the range where $J_{\text{shift}} \propto V$ and second, $\Delta E_F(V) \propto V$ has to hold, i.e. the quasi-Fermi level splitting has to change linearly with the external voltage.

We can compare these findings with the data shown in Figure 7b. The value of $\Delta E_F(V)$ changes approximately linearly with external voltage until about 0.9 V. However, there is a small superlinear contribution to the rise that can also be seen in double-logarithmic version of Figure 7b that is found in the supporting information (Figure S6). Furthermore, we note that the slope of the thin solid lines in Figure 7b is only 4 meV/V. Thus, within the linear region, $\Delta E_F(V)$ increases by <4 meV which is much less than $n_{\text{id}}kT \approx 50$ meV. Thus, we note that the for the whole region of the shifted current until the transition to the exponential rise of the $J_{\text{shift}}$ vs. $V$, the Taylor expansion of equation 17 is dominated by the linear terms. The second condition, namely $\Delta E_F(V) \propto V$ does not hold perfectly. Instead, the increase is slightly superlinear but not at all exponential up to about the voltage $V_{\text{mpp}}$ at the maximum power point.

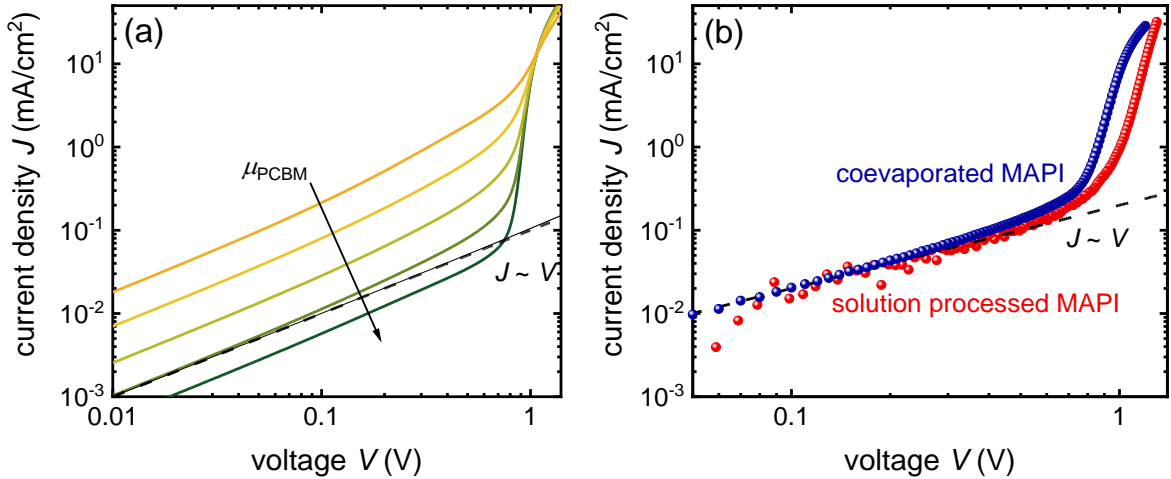

**Figure 9:** (a) Simulated shifted current densities $J_{shift}$ as presented in Figure 5d plotted double-logarithmically. A curve with slope 1, i.e. $J \sim V$ is shown as a dashed line. (b) Experimental data of $J_{shift}$ vs. V for the solar cells shown in Figure 2 (coevaporated MAPI cell) and the cell used for Figures 7 and 8 (solution-processed MAPI). The deviations from linearity originate from $\Delta E_F(V)$ being slightly non-linear with $V$.

Hence, the shifted currents can already reveal certain non-trivial features of the voltage-dependent photoluminescence that allow us to draw conclusions on the band diagram of the solar cell. As seen in Figure 5, the linearity of $J_{shift}$ and $V$ over a large voltage range is a feature that only exists in the presence of a screened electric field in the absorber. In the presence of a substantial electric field in the absorber (as would be the case for the scenario shown in Figure 5b), the higher order terms of the Taylor-expansion would matter already at fairly low voltages. Unfortunately, we cannot make any strong statement about the magnitude of $J_{rec}(0)$ from the shifted current alone. According to equation 17, the shifted current will be equal to $J_{rec}(0)$, exactly if $\Delta E_F(V) - \Delta E_F(0) = \ln(2) n_{id} kT$ which corresponds to 36 meV for $n_{id} = 2$. This value is substantially above the linear region, which implies that the recombination current at short circuit must be higher than $J_{shift}$ at the voltage where it becomes visibly non-linear with voltage. Thus, when observing an approximately linear region up to a current of 1 mA/cm² as seen in Figure 8a, we can already conclude that significant recombination losses at short circuit exist even before we measure the voltage dependent photoluminescence.

Note that the concept of the shifted currents also lends itself to an analysis of literature data. Here, it is decisive to obtain measured data with a significant precision rather than picking data from the pdfs of publications. In the supporting information, we show some published data from recent high efficiency solar cells that we replotted in the style of Figure 2. This data includes the 86% fill factor solar cell from Peng and coworkers[26] and the 24% efficient solar cell from Tan et al. [74] All of the data shows significant, weakly voltage dependent recombination currents at small forward bias that appear like a photoshunt

## 8. Conclusions

We discuss two approaches to shed light on recombination losses in perovskite solar cells that occur at voltages below $V_{oc}$, i.e. in situations where the efficiency of charge carrier transport through both the perovskite as well as the charge transport layers affects the rate of recombination. In this situation, the recombination losses will be visible as a reduced fill factor and/or reduced short-circuit current density. All approaches shown in the paper share the common idea that deviations from the simple superposition principle, i.e. the ability to sum up dark current and photocurrent to obtain the current under illumination, are exploited to identify and quantify losses. All of these losses are recombination losses that have a resistive component, i.e. they would disappear if the conductivities of all layers within the solar cell were infinitely high. Thus, the methods all exploit the fact that $\Delta E_\text{F}(V) \neq qV$ is typically valid for all bias situations other than open circuit. Given that the losses we are interested in are both related to recombination and resistive effects, we can approach the task of quantifying these losses both from the perspective of methods aimed at quantifying resistive losses and methods aimed at characterizing recombination losses.

The first approach is based on the comparison of light and dark current-voltage curves to determine the series resistance. This method is a standard method in crystalline silicon solar cells but has so far been rarely used for thin-film solar cells. We show that we obtain a highly non-linear (i.e. voltage dependent) series resistance. Using numerical simulations, we show that the behavior is qualitatively consistent with a weakly conductive, intrinsic charge transport layer (either ETL or HTL) and a relatively field-free perovskite absorber. To arrive at a more complete picture of voltage dependent recombination, we use a second approach that is based on voltage dependent photoluminescence. Unlike in previous approaches, we take care to minimize any effect on the data that is due to uncontacted perovskite by removing it with a laser. We then use a simple model of recombination inside a diode with $\Delta E_F(V) \neq qV$ to arrive at a quantitative estimate of the recombination current as a function of voltage. Furthermore, we deduce from the application of both methods that the electric field in the perovskite absorber must be largely screened to be consistent with the data. We therefore confirm earlier conclusions on field screening[61, 75, 76] using a very simple approach that only requires the illuminated current-voltage curves.

**Acknowledgements**

The authors thank Rene Krimmel and Lisa Krückemeier for the fabrication of the coevaporated solar cell used for Figure 2. The authors acknowledge financial support from the Helmholtz Association via the project PEROSEED and via the project-oriented funding (POF IV).